\documentclass[12pt]{iopart}

\usepackage{graphicx}  
\begin{document}

\title{Limits on coupling between dark components}

\author{ Roberto Mainini, Silvio Bonometto}

\address{Department of Physics G.~Occhialini -- Milano--Bicocca
 University, Piazza della Scienza 3, 20126 Milano, Italy \&
I.N.F.N., Sezione di Milano}

\begin{abstract}
DM--DE coupling can be a phenomenological indication of a common
origin of the dark cosmic components. In this work we outline a new
constraint to coupled--DE models: the coupling can partially or
totally suppress the Meszaros effect, yielding transfered spectra with
quite a soft bending above $k_{hor,eq}.$ Models affected by this
anomaly do not show major variation in the CMB anisotropy spectrum and
it is herefore hard to reconcile them with both CMB and deep sample
data, through the same value of the primeval spectral index.

\end{abstract}

\pacs{98.80.-k, 98.65.-r }
\maketitle
\section{Introduction}
\label{sec:intro}
There can be little doubts that a tenable cosmological model must
include at least two dark components, cold Dark Matter (DM) and Dark
Energy (DE); yet only hypotheses on their nature exist, most of them
assuming that DM and DE are physically unrelated and that their
similar densities, in today's world and just in it, are purely
accidental.

Attempts to overcome this conceptual deadlock were made by several
authors suggesting, first of all, that DE has a dynamical nature
\cite{DDE} (for a review see \cite{PR} and references therein).  An
alternative idea is that DE is a phenomenological consequence of the
emergence of nonlinearity; this appealing option was repeatedly
considered (see, e.g., \cite{Kolb} and references therein), but is far
from being shown and leaves however apart the question of DM nature.
Interactions between DM and dynamical DE \cite{interaction} (see also
\cite{interaction2}) might partially cure the problem, keeping close
values for their densities up to large redshift. This option could
also be read as an approach to a deeper reality, whose physical
features could emerge from phenomenological limits to coupling
strength and shape.

A longer step forward was attempted by \cite{Mainini2004}, suggesting
that DM and DE derive from a single complex scalar field, being its
quantized phase and modulus, respectively. The complex field could be
the one responsible for $\cal CP$ conservation in strong interactions,
within a scheme similar to Peccei \& Quinn framework \cite{PQ} (see
also \cite{axion}). At variance from previous suggestions, which
introduce parameters and aim at limiting them through data fitting,
this option --$\, $dubbed {\it dual--axion} model$\, $-- cuts the
available degrees of freedom, including as many parameters as a
standard--CDM approach. It is then quite appealing that its reduced
parameter budget is sufficient to fit quite a number of observational
constraints \cite{cmb}, still allowing for a common nature of DM and
DE and for a specific shape of interaction between them.

In this paper, however, we keep on the phenomenological side and
discuss generic constraints to DM--DE interactions. This discussion
will have a fallout also on the {\it dual--axion} approach, which does face a
problem, because of the feature of the DM--DE coupling it causes.

A coupling of baryons with DE is ruled out by observational
consequences similar to modifying gravity. Limits are looser for
DM--DE coupling, whose consequences can be appreciated only over
cosmological distances. It should also be outlined that forces acting
within the dark sector could modify predictions on high concentration
DM lumps. There can be little doubts that cold DM particles, feeling
gravity only, give them NFW profiles. Yet, observational data do not
lend much support to this shape, for any scale range, and direct
interaction between DM particles is severely constrained also by
recent data.

This is a further reason to consider DM--DE interactions, either as a
fundamental theory or as an effective framework to approach deeper
physics. It is then important to devise any observational limit to
such interactions and, in this paper, we outline further constraints
to its shape; they are consequences of the early behavior of density
fluctuations, over scales destined to evolve into non--linear structures.

Fluctuations over such scales enter the horizon before
matter--radiation equality and their growth is initially inhibited by
the overwhelming density of the radiative component, then still
behaving as a single fluid together with baryons. While fluctuations
in the fluid behave as sonic waves, self gravitation of DM is just a
minor dynamical effect is respect to cosmic expansion. This freezing
of fluctuation amplitudes until equality is known as {\it Meszaros
effect}.

The main point we wish to outline here is that DM--DE coupling can
damp Meszaros effect, so that the rate of fluctuation growth, between
the entry in the horizon and equality, is significantly enhanced. As
a matter of fact, fluctuation freezing is essential, in shaping the
transfered spectrum, which peaks on the scale $k_{hor,eq}$ entering the
horizon at equality. At smaller mass scales ($k>k_{hor,eq}$) the spectrum
declines because of the increasing duration of the freeze.

The freezing or its damping have modest consequences on the evolution
of fluctuations in baryons and radiation, evolving then in the sonic
regime. What we shall therefore find are significant changes in the
transfer funcions, while CMB spectra keep almost unaffected.

Constraints to coupled DE models arise from both linear and
non--linear effects. It has been known since long that coupling may
cause a $\phi$--MD epoch after matter--radiation equality (see, e.g.,
\cite{constraints}). This changes the (comoving) distance of the last
scattering band. In order to fit data, the present value of the Hubble
parameter $H_o$ needs then to be increased. Limits on $H_o$ turn then
into limits to the coupling.

%on-to
Limits to the coupling, in the case of a Ratra Peebles \cite{RP}
self--interaction potential, where also found in \cite{maccio}, by
studying halo concentration distibution.

%constraint-feature
The feature outlined in this work affects the transfer function,
leaving almost unaffected CMB anisotropies. Discussing how the
transfer function is affected by DM--DE coupling is the main aim of
this technical paper. We shall also exhibit CMB angular spectra, to
confirm that they suffer just marginal changes.

%soppresso ``We warn the reader''
No general data fitting, constraining parameters and/or showing
specific model advantages, will be made here. In fact, what we wish to
outline is a major effect, which allows to discard a class of models,
{\it a priori}. This is why we keep to cosmological parameter values
ensuing from WMAP3 best--fit \cite{Spergel2006}, although deduced by
assuming a $\Lambda$CDM model. In particular, we shall take an overall
density parameter $\Omega = 1$; the present value of the cold DM
(baryon) density parameter will be $\Omega_{o,c} = 0.224$
($\Omega_{o,b} = 0.044$); the dimensionless Hubble parameter will be
$h = 0.704;$ the primeval spectral index, when not taken as a free
parameter, will be $n = 0.947\, $.

Within this frame we shall consider a self--interacting scalar field,
causing cosmic acceleration when its pressure/density ratio $w =
p_{DE}/\rho_{DE} $ falls in the range $(-1,-1/3).$ Quite in general,
it is
\begin{equation}
\rho_{DE}=\rho_{k,DE}+\rho_{p,DE} \equiv {{\dot{\phi }}^{2}/2a^{2}}+V(\phi ),~
~~~~
p_{DE}=\rho_{k,DE}-\rho_{p,DE}~,
\end{equation}
so that it is
$-1/3 \gg w>-1$ when dynamical equations yield $\rho _{k,DE}/V\ll 1/2$. Here
\begin{equation}
ds^{2} = g_{\mu\nu} dx^\mu dx^\nu =
a^{2}(\tau) (-d\tau^{2}+ dx_i dx^i ) ~~~~~~ (i=1,..,3)
\label{metric}
\end{equation}
is the background metric and dots indicate differentiation with
respect to $\tau $ (conformal time). The $w$ ratio exhibits a time
dependence set by the shape of $V(\phi)$. Much work has been done on
dynamical DE (see, e.g., \cite{PR} and references therein), also
aiming at restricting the range of acceptable $w(\tau)$'s, so gaining
an observational insight onto the physics responsible for the
potential $V(\phi)$.

Our analysis here will however be restricted to SUGRA 
potentials \cite{Brax}
\begin{equation}
\label{ptntl}
V(\phi) = (\Lambda^{\alpha+4}/\phi^\alpha) \exp(4 \pi \phi^2/m_p^2)
\end{equation}
admitting tracker solutions. 
Here $m_p = G^{-1/2}$ is the Planck mass.  
This will enable us to focus on peculiarities caused by the
coupling. Let us also remind that, once the DE density parameters
$\Omega_{DE}$ is assigned, either $\alpha$ or the energy scale
$\Lambda$, in the potentials (\ref{ptntl}), can still be freely
chosen. In this paper we show results for $\Lambda = 10^2\, $GeV.

The SUGRA potential, at least in the absence of coupling, yields an
excellent fit of observational data \cite{gerva}. We tested the
effects of coupling for a number of values of the scale $\Lambda$,
from 10 to $10^4$, and also changing the shape of the potential into
Ratra--Peebles.

In the former case we find just marginal shifts. In the latter one,
quantitative changes can be significant. The overall behavior is
however identical and this potential is known to yield a poor fit to
CMB data, unless quite a small $\Lambda$ scale is taken, so spoiling
its physical appeal.

Within this frame we aim at focusing problems and showing the
quantitative consequences of different options.

\section{Dynamical equations}
Let us then start from the background equations for DE and DM
when the metric is (\ref{metric}), reading
\begin{equation}
\ddot{\phi} + 2{\dot a \over a} \dot{\phi} + a^2 V_{,\phi} = 
C(\phi) a^{2} \rho _{c}~, ~~~~ \dot{\rho _{c}} + 3 {\dot
a \over a}\rho_{c} = - 
C(\phi) \dot \phi \rho _{c}~
\label{backg}
\end{equation}
where we set
\begin{equation}
C(\phi) = 4 \sqrt{\pi \over 3} {\beta\over m_p} 
\left( \phi \over m_p \right)^\epsilon
= 4 \sqrt{\pi \over 3} {\tilde \beta\over m_p} 
%= 4 \sqrt{\pi \over 3} {\beta \over m_p} 
\label{C}
\end{equation}
%Here $m_p = G^{-1/2}$ is the Planck mass.  
A possible time dependence
of the coupling strength was considered but not deepened since the
early work of \cite{Amendola04}. The {\it dual--axion} model naturally
predicts a coupling $C=1/\phi$, consistent with eq.~(\ref{C}), if
$\epsilon = -1$ and $\beta \simeq 0.244. $ Quite in general, for
dimensional reasons, $C$ can be expressed through products of $m_p^a$
and $\phi^b$ with $a+b = -1.$ During most cosmic evolution $\phi$ is a
monotonically increasing function, so that the sign of the exponent of
$\phi$ tells us whether the coupling was stronger of weaker in the
past. (An exception can be recent times, as\ the exponential term in
the SUGRA potential can cause a re-bounce of $\phi$ when it approaches
$m_p;$ our arguments here concern much earlier times, when it is
safely $\phi < m_p$).

An expression of $C,$ made by a polynomial including terms with
different powers of $\phi,$ could select a peculiar epoch to have then a
weaker or stronger coupling. Such an option, however, is clearly {\it
ad--hoc} and does not seem to deserve further investigation.

An explicit dependence of $C$ upon time would be hard to
reconcile with the Lorentz invariance of the Lagrangian mass term
\begin{equation}
{\cal L}_c = - B(\phi) m_\chi \bar \chi \chi
\end{equation}
setting the coupling between the DE scalar field $\phi$ and a spinor
field $\chi$ supposed to yield DM, as $C(\phi) = d (\ln B)/d\phi.$
(Notice that $B(\phi) m_\chi$ is the time--dependent mass of DM
quanta). A dependence of $C$ on $\phi$ seems therefore the only
way to instaure a time--dependent coupling.

Let us then describe fluctuation equations. In the period when
Meszaros effect occurs, DM, photons and baryons can be treated as
fluids; (massless) neutrinos, instead, are not a fluid.  Fluctuations
in a generic fluid with $p/\rho = w$ and $\delta p/\delta \rho =
c^2_s$ fulfill the equations
%$$
\begin{eqnarray}
\fl
\dot \delta = - (1+w) (kv + \dot h/2) - 3 (c_s^2-w)({\cal H}-C \dot \phi)
\delta -  (1-3w)  ( C \dot \varphi  +
C' \dot \phi \varphi)
\nonumber
%$$
\\
\fl
%\begin{equation}
\dot v = - (1-3w) ({\cal H} - C \dot \phi) v + {c_s^2 \over 1+w} k \delta - {\dot w
\over 1+w} v - k \sigma  -k C {1-3w \over 1+w} \varphi 
~.
\label{fluids}
\end{eqnarray}
Here $\delta = \delta \rho/\rho$, $v = i u_i k^i/k$, $(\rho + p)
\sigma = -(\hat k_i \cdot \hat k_j - \delta_{ij})(T^{ij}-\delta_{ij}
T^k_k /3)$ ($u_i$ are the space components of the velocity field in
the fluid and $T^{ij}$ is its stress--energy tensor) and ${\cal H} = \dot a /a$,  while
the DE field
\begin{equation}
\phi (\tau,{\bf x}) = \phi_o (\tau) + \varphi (\tau,{\bf x})
\end{equation}
is split into a background component $\phi_o$, coinciding with the
$\phi$ field obeying the eq.~(\ref{backg}), and the space--dependent
fluctuation $\varphi;$ in eqs.~(\ref{fluids}), Fourier components
of $\varphi$ are considered which fulfill the equation:
\begin{equation}
\ddot \varphi + 2{\cal H} \dot \varphi + k^2 \varphi 
+ a^2 V''_\phi \varphi + \dot \phi \dot h/2 =
Ca^2 \rho_c \delta_c + C'_\phi a^2 \rho_c \varphi~,
\label{kuku}
\end{equation}
while fluctuation self--gravity is fully accounted by $\dot h$,
obtained by integrating the equation
\begin{equation}
\label{ddoth}
\ddot h + {\cal H} \dot h = - 8\pi G \left\{ a^2 \rho [\Omega_{\gamma
b} (1+3c_s^2) \delta_{\gamma b} + \Omega_c \delta_c] - 2a^2 V'_\phi \varphi +
\dot \phi \dot \varphi \right\}
\end{equation}

In the absence of coupling ($C = 0$), we can take $v=0$ in
eqs.~(\ref{fluids}) without loss of precision, and face the dynamical
problem though a single first order equation.  This is no longer true
when DM particles can be pushed by DE forces. Then the DM equation, on
scales below the horizon and on times before matter--radiation
equality, reads
\begin{equation}
\fl
\ddot \delta_c + \left[ {\dot a \over a} -  {4 \over m_p} \sqrt {\pi \over 3} 
 \tilde
\beta(\phi) \dot \phi \right] \dot \delta_c - 4\pi G a^2 \rho \left\{
\left[1+ {4 \over 3} \tilde \beta^2(\phi) \right] \Omega_c \delta_c + 
 \Omega_{\gamma b}
\delta_{\gamma b} \left[1+3 c_s^2 \right] \right\} = 0~,
\label{m0}
\end{equation}
Here, $\rho$ is the overall density; $\Omega_c$ and $\Omega_{\gamma
b}$ are time dependent density parameters for DM and the
baryon--photon fluid. To show eq.~(\ref{m0}) one needs a little
algebra, which will be postponed to the end of the section.

Still from eqs.~(\ref{fluids}), we can also work out the equations for
the photon--baryon fluid, by taking $C=0$ and 
\begin{equation}
c_s^2 \equiv \delta p_{\gamma} / (\delta \rho_b + \delta \rho_{\gamma} )
= \left[3 (1+3\Omega_b /4 \Omega_{\gamma})\right]^{-1}~,
\end{equation}
\begin{equation} 
w_{\gamma b}(a) \equiv p_{\gamma} / (\rho_b + \rho_{\gamma} ) =
\left[3 (1+\Omega_b / \Omega_{\gamma})\right]^{-1}~.
\label{wgammab}
\end{equation}
Here $\Omega_b$ and $\Omega_\gamma$ are the time dependent baryon and
photon density parameters.  The baryon component can be responsible
for a shift of $w_{\gamma b}$ from 1/3; although initially small, it
can approach 20--25$\, \%$ at the eve of recombination. Notice then
that a non--vanishing factor $1-3w \sim \Omega_b/\Omega_\gamma$ keeps
a direct influence of $\varphi$ on baryon--photon fluctuations.

This set of equations enables the reader to build a simplified
numerical algorithm, directly testing the suppression of Meszaros'
effect.

Clearly, to study the later evolution, since the eve of baryon--photon
decoupling a full kinetic treatment of the radiation is needed. This
will be used to confirm that CMB anisotropies are just marginally
affected, but is unessential to focus on the greater changes occurring
to the transfer function.

Let us now summarize the procedure to obtain eq.~(\ref{m0}); a reader
unintersted in it can skip the rest of this section.

The starting point are again the eqs.~(\ref{fluids}). Together with
them, let us consider again eq.~(\ref{kuku}). There, any mass--like
term multiplying $\phi,$ in comparison with $k^2 ,$ is negligible;
then, before equality, it yields
\begin{equation}
\ddot \varphi + (2 / \tau) \dot \varphi  + k^2
\varphi = Ca^2 \rho_c \delta_c - \dot \phi \dot h/2
\label{kvkv}
\end{equation}
Here, two kinds of time dependence must be compared, over fluctuation
and Hubble time scales. 
%Terms exhibiting just the latter dependence
%shall be considered {\it constant}, in comparison with fluctuating terms. 
Accordingly, we can express $\varphi$ as sum of rapidly and slowly
varying terms by actually summing up the (rapidly varying) general
integral of the equation obtainable by equating to zero the l.h.s.
and a (slowly varying) integral obtained by equating the last term at
the l.h.s. with the r.h.s.$\, $. If we then time--average over
fluctuation time scales, only the latter contribution survives and
\begin{equation}
\label{phix}
\langle \varphi \rangle \simeq 
{1 \over k^2} \left(Ca^2 \rho_c \delta_c - \dot \phi \dot h/2 \right)
\end{equation}
Let us recall that dropping the contribution of fluctuating terms is
the standard procedure to obtain an analytical description of
Meszaros' effect. The point here is that, while all $\varphi$
derivatives can be dropped, there is a slowly varying contribution to
$\varphi$ which cannot be soon disregarded.

However, if the expression (\ref{phix}) is used to replace $\varphi$
in eqs.~(\ref{fluids}) for CDM, yielding
\begin{equation}
\label{neglect}
\dot \delta_c + k v_c + {\dot h \over 2} =
%- {1 \over k^2} {d^2 \over d\tau^2} \left(C^2 a^2 \rho_c \delta_c \right) = 
- 2 \beta^2  {d \over d\tau} \left[ \left( \phi \over
  m_p \right)^{2\epsilon} { \Omega_c \delta_c \over (k\tau)^2 } \right]
+ 4 \sqrt{\pi \over 3} {\beta \over k^2} {d \over d\tau} \left(\dot \phi \dot h
\over \phi^{-\epsilon} m_p^{1+\epsilon} \right)
\end{equation}
it becomes clear that also the slowly varying $\varphi$ contributions
can be dropped. In fact, the first term at the r.h.s.~contains a
division by $(k\tau)^2,$ which is the squared ratio between long and
short timescales and, altogether, it is $\sim {\cal O}[\dot \delta /
(k\tau)^2] \ll \dot \delta .$ We must then acknowledge that a time
derivative of a quantity $Q$, varying over the Hubble time scale, is
$\sim {\cal O}(Q/\tau) .$ The second term at the r.h.s. is then $\sim
{\cal O} [(\phi/m_p)^{1+\epsilon} \dot h/(k\tau)^2],$ while we took $1 +
\epsilon > 0$ and, in the epoch considered, $\phi/m_p \ll 1.$ It must
then be even smaller than the the first one.

Setting then to zero the l.h.s. of eq.~(\ref{neglect}), we obtain the
relations
\begin{equation}
\ddot \delta_c + k\dot v_c + {\ddot h \over 2} \simeq 0~,~~~~~~
 k v_c \simeq -\dot \delta_c - {\dot h \over 2} ~.
\label{two}
\end{equation}
In turn, the second eq.~(\ref{fluids}), using this latter equality,
yields the relation
\begin{equation}
k \dot v_v = (C\dot \phi -{\cal H}) \left[ -\dot \delta_c - {\dot h/ 2} \right]
-k^2C\varphi ~,
\end{equation}
which can be replaced in the former eq.~(\ref{two}), together with
eq.~(\ref{ddoth}) and (\ref{phix}), obtaining
$$
\ddot \delta_c + \left[ {\cal H} - C \dot \phi \right] \dot \delta_c %-
%$$
%\begin{equation}
 - 4\pi G \left\{ a^2 \rho [\Omega_{\gamma b} (1+3c_s^2)
\delta_{\gamma b} + \Omega_c \delta_c] - 2a^2 V'_\phi \varphi + \dot
\phi \dot \varphi \right\} - C^2 a^2 \rho_c \delta_c = 0
%\end{equation}
$$ 
Neglecting here $\varphi$ fluctuations as source terms --$\, $DE
yields a minor contribution to the overall density, as shown in
Fig.~\ref{omega}$\, $-- and using eq.~(\ref{C}),
eq.~(\ref{m0}) is soon obtainable. 

\section{Overcoming the freeze}
Let us then focus our attention on eq.~(\ref{m0}) and outline first
that, in average, the term $\Omega_{ \gamma b} \delta_{\gamma b}
\left(1+3 c_s^2 \right)$ almost vanishes, as sonic waves
fluctuate. Then, in the absence of coupling ($\tilde \beta=0$), the
self--gravitation term $\propto \delta_c$ is also damped by $\Omega_c
\ll 1.$ If the photon--baryon term is then neglected, the increasing
mode (approximately) reads
\begin{equation}
\delta_c \propto 1 + 3y/2
\end{equation}
with $y = 3 w_{\gamma b}\, a/a_{eq}$ (see, e.g., \cite{ColLuc}) and this
yields a growth from horizon to equality never exceeding a factor
2.5$\, $, that we shall approximate as $\delta_c \propto a^{1/4},$
according to numerical outputs. Meanwhile, above the horizon, in a
synchronous gauge, $\delta_c \propto a^2\, $. Hence, for $k > k_{hor,eq},$
the growth is slowed down by a factor $\propto a_h^{7/4}(k),$ 
while the scale factor when $k$ passes through the horizon, $a_h(k) \propto
k^{-1}.$ Altogether, for $k > k_{hor,eq},$ we expect a transfered spectrum
$P(k) = A \, k^{n} {\cal T}^2(k) \simeq A k^{n - 3.5}$ (${\cal T}(k)$
is the transfer function).

%%%%%%%%%%%%%%%%%%%%%%%%%%%%%%%%%      
\begin{figure}
\begin{center}
\includegraphics*[width=12cm]{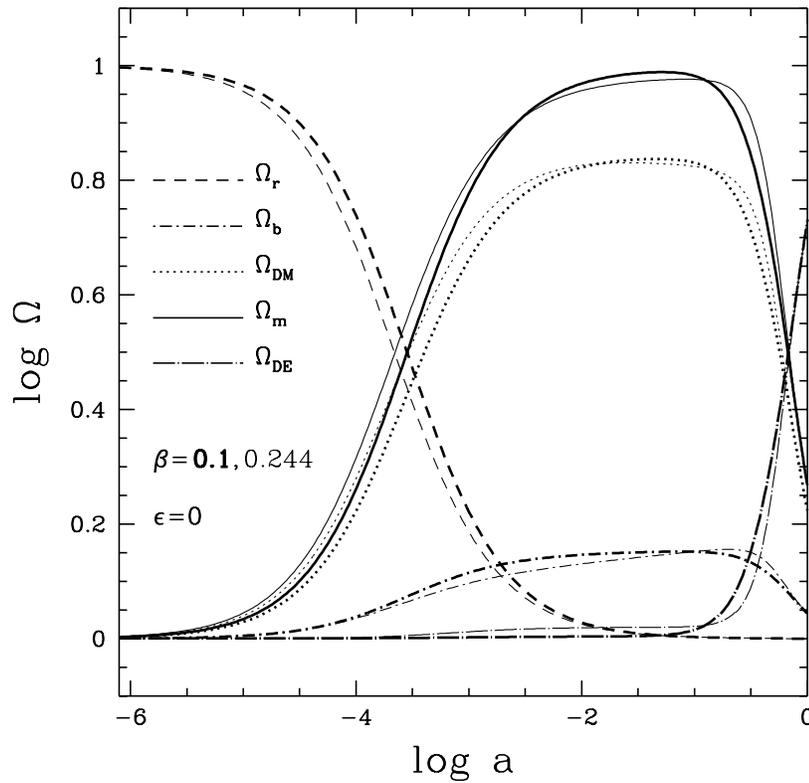}
\end{center}
\vskip -1.5truecm
\caption{Scale dependence of the density parameters of the
various components in coupled DE models with constant coupling.
This plot shows also the displacement of $z_{eq}$ and, henceforth, 
of $k_{hor,eq}$ as $\beta$ increases: thicker (thinner) curves refer
to $\beta = 0.01$ (0.0244). }
\label{psplit}
\vskip -.04truecm
\end{figure}
%%%%%%%%%%%%%%%%%%%%%%%%%%%%%%%%%
%%%%%%%%%%%%%%%%%%%%%%%%%%%%%%%%%      
\begin{figure}
\vskip -.8truecm
\begin{center}
\includegraphics*[width=11cm]{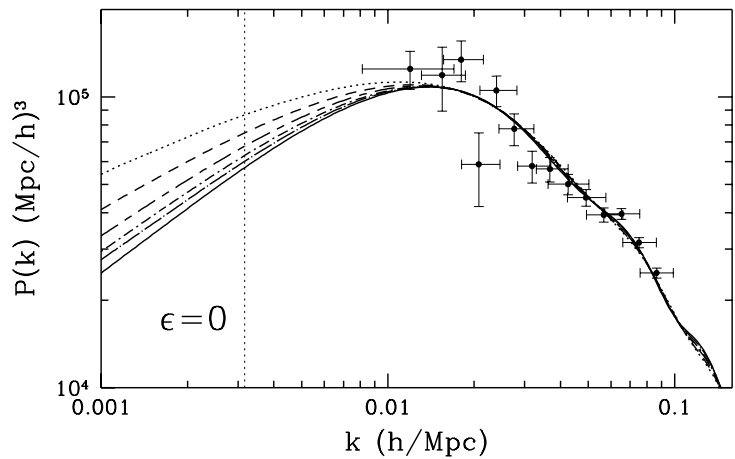}
\end{center}
\vskip -4.truecm
\caption{Best fits of SDSS data for constant $\beta$ from 0 (solid
line) to 0.25 (dotted line). Different lines correspond to a $\beta$
increase by 0.05$\, $. The vertical dotted line yields the scale of
$C_{10}.$ }
\label{fitbeta0}
%\vskip -.4truecm
\end{figure}
%%%%%%%%%%%%%%%%%%%%%%%%%%%%%%%%%
%%%%%%%%%%%%%%%%%%%%%%%%%%%%%%%%%      
\begin{figure}
\begin{center}
\includegraphics*[width=11cm]{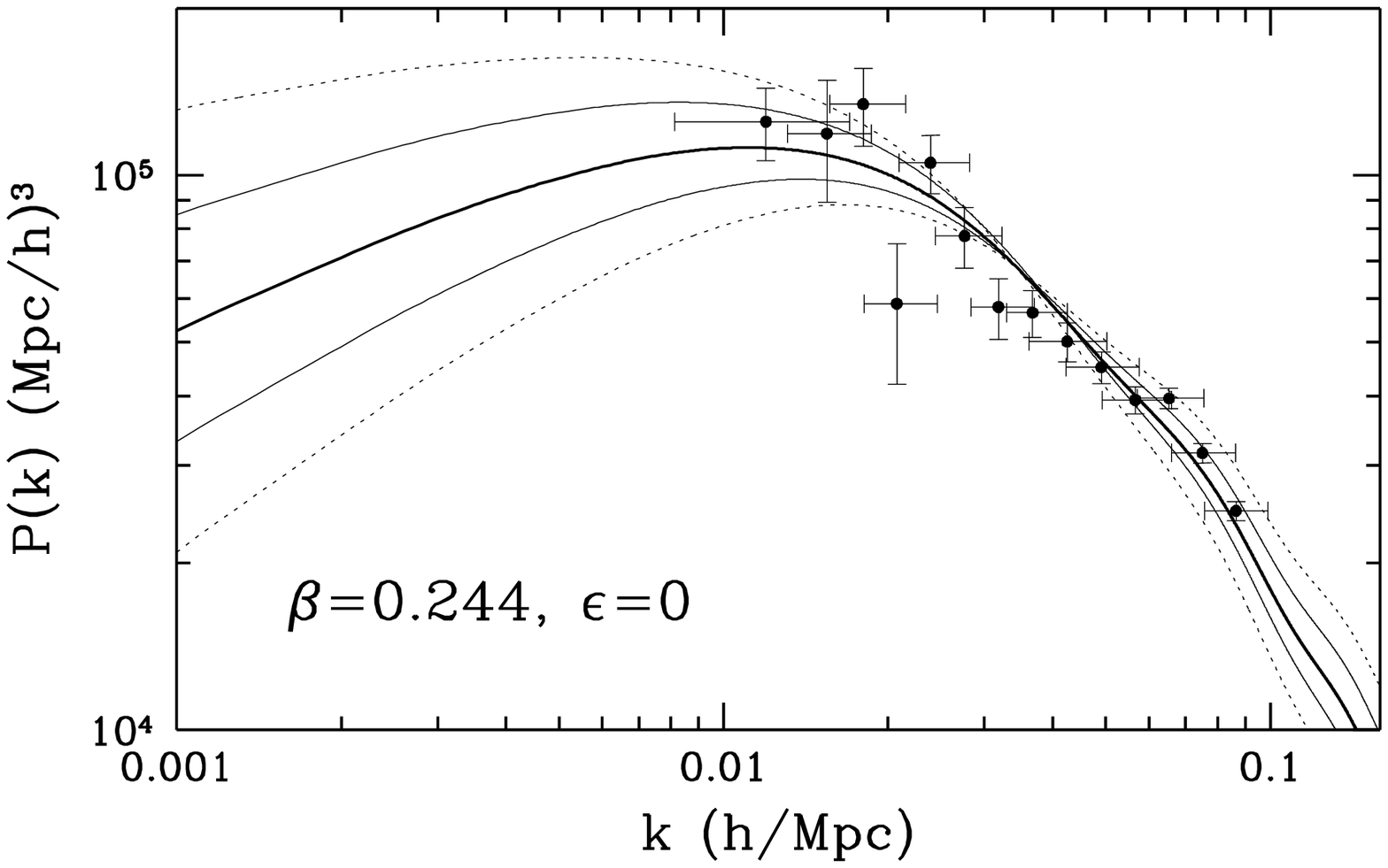}
\end{center}
\vskip -4.5truecm
\caption{If values of $n$ at 1-- or 2--$\sigma$'s from best fits
are taken, spectra are significantly modified. Here we show the
effect in the case with $C=1/m_p.$   }
\label{fitsigma2}
%\vskip -.4truecm
\end{figure}
%%%%%%%%%%%%%%%%%%%%%%%%%%%%%%%%%
%%%%%%%%%%%%%%%%%%%%%%%%%%%%%%%%%      
\begin{figure}
\begin{center}
\includegraphics*[width=11cm]{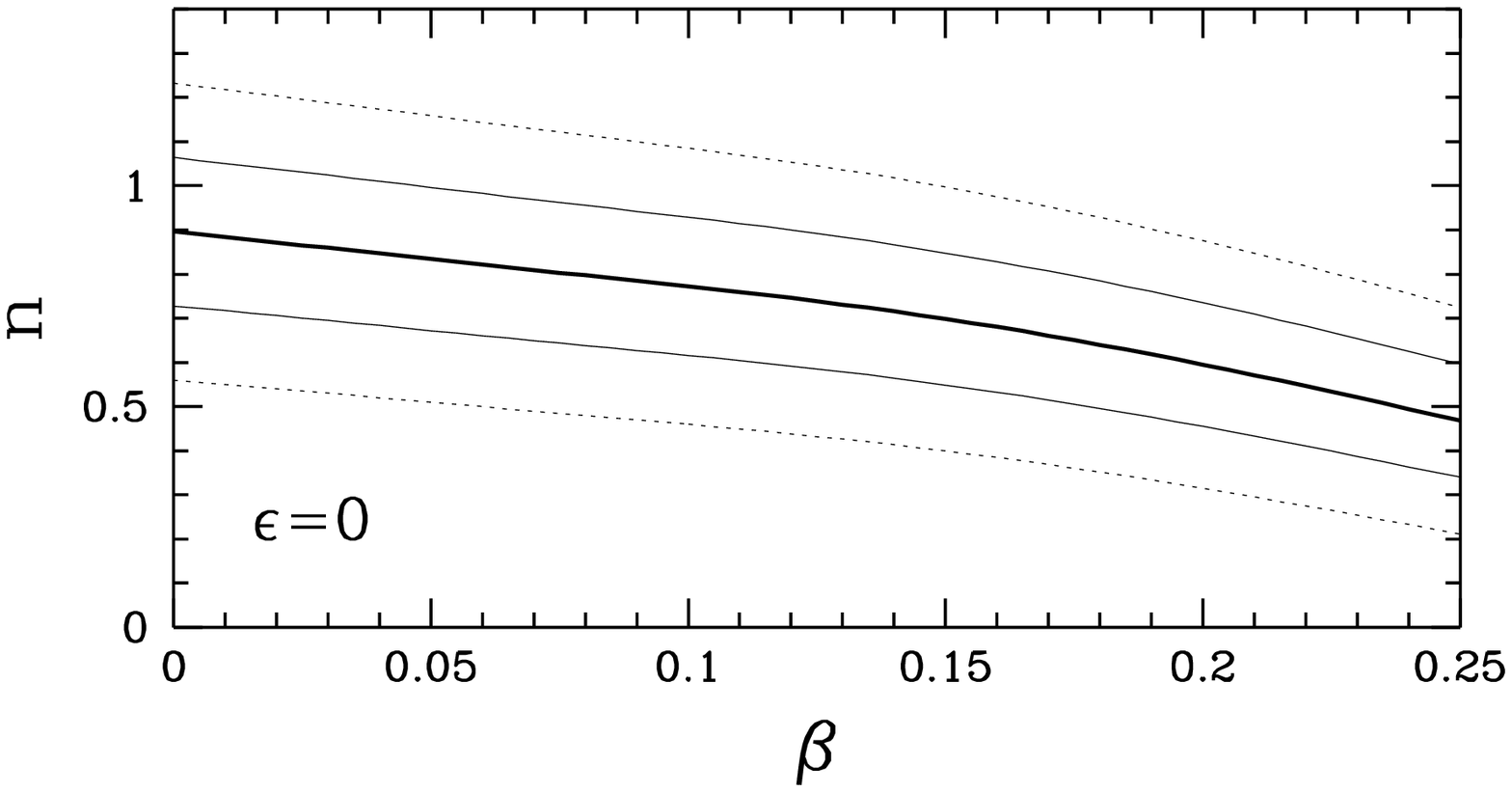}
\end{center}
\vskip -5.9truecm
\caption{$n$ intervals for increasing (constant) coupling strength }
\label{ennebeta}
%\vskip -.4truecm
\end{figure}
%%%%%%%%%%%%%%%%%%%%%%%%%%%%%%%%%

It is then easy to see what can change because of the coupling.  The
coefficient of the {\it friction term} $\propto \dot \delta_c$ is
certainly reduced and can even invert its sign. Even more
significantly, the term $[1+ 4 \tilde \beta^2(\phi)/3] \Omega_c $ can
attain or overcome unity, not only because of the greater size of
$\tilde \beta (\phi),$ but also thanks to the modified time dependence
of $\Omega_c.$ It turns out, in fact, that a decreasing dependence of
$\tilde \beta$ on $\phi$ ($\epsilon < 0$) yields higher $\Omega_c$
values in the relevant redshift interval. In general, $\phi$ is
smaller at earlier times and $\epsilon < 0$ causes a stronger coupling
in the past.

Let us however debate first the constant coupling case ($\epsilon =
0$).  Available data set then a constraint $\beta < 0.1$--0.2$\, $
\cite{constraints}, \cite{maccio} (see also \cite{constraints2}; beware of the
different coupling definition), limiting the acceptable discrepancy of
DM and DE evolution, after recombination, from uncoupled models. The
most direct effect, in this case, concerns large scales entering the
horizon late, when such discrepancies occur. The low--$l$ plateau
of the CMB anisotropy spectrum can then undergo a modified ISW effect,
while some changes in the $C_l$ behavior, up to the first peak at
$l \sim 200,$ can be compensated by slightly modifying $n$, $h$ and
other parameters.

But the most significant shifts occur on the transfer function. Its
slope, at $k > k_{hor,eq}$ and up to a scale $k \sim 0.1/{h{\rm
Mpc}^{-1}},$ where non--linearity effects become important, is
slightly distorted as $\beta$ increases. The main effect, however, is
a progressive displacement of $k_{hor,eq}$ itself. In Figure \ref{psplit}
we compare the $\Omega$ evolution in models with different constant
coupling, showing the significant displacement of the crossing between
$\Omega_r$ and $\Omega_c$, when different $\beta$'s are taken.

We can show the impact of this displacement by fitting transfered
spectra, over these scales, with the Luminous Red Galaxies sample data
from the Sloan Digital Sky Survey (SDSS) \cite{Tegmark} and allowing
the primeval spectral index $n,$ assumed to be constant, to act as a
free parameter.

In Figure \ref{fitbeta0} we show the result of this fit.  Transfered
spectra, when $k_{hor,eq}$ vary, easily accommodate deep sample data in the
linear range, as the $k_{hor,eq}$ shift is compensated by a slightly
smoother slope. All that however requires a non negligible decrease of
$n,$ and transfered spectra risk to become too high in the region
where they should fit CMB data (the scale of the 10--pole is indicated
in the Figure).

An attempt to balance this spectral distortion can be made by varying
other parameters. It is then significant to consider Figure
\ref{fitsigma2}, showing deep sample data vs.~transfered spectra
computed with $n$'s within 1-- and 2--$\sigma$'s from the
best--fit. In the Figure we took $\beta = \sqrt{3/\pi}/4 \simeq
0.244$. Then, in Figure \ref{ennebeta}, we show the 1-- and
2--$\sigma$ range of $n,$ for constant coupling strength. Let us also
recall that likelihood analysis showed that models with constant
$\beta < \sim 0.2$ do not exhibit severe disagreements with data.

\begin{figure}
\begin{center}
\includegraphics*[width=15cm]{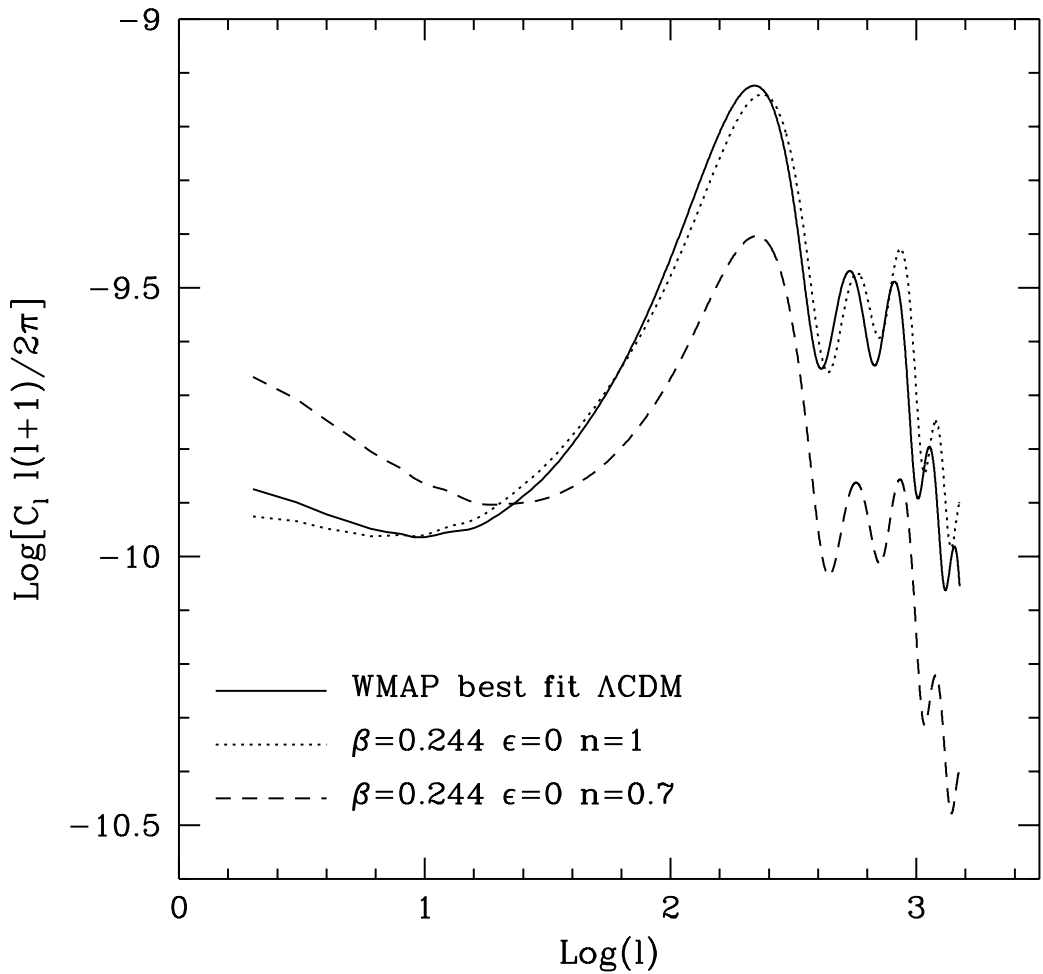}
\end{center}
\vskip -1.2truecm
\caption{Anisotropy spectrum of the $\Lambda$CDM model yielding the
best fit to WMAP3 data compared with the spectra for coupled models
with $\beta = 0.1$ and $\epsilon = 0,$ for $n = 1 $ and $n = 0.7.$
Already in the former case some difference exists, but no major
qualitative changes occur; by adjusting other model parameters one can
expect a reasonable fit to data. Taking $n = 0.7,$ any fitting to CMB
anisotropy data is apparently excluded.  }
\label{Cl0}
%\vskip -.4truecm
\end{figure}

The effect on CMB data fitting arising from $n$ values distant from
unity can be also directly inspected in Figure \ref{Cl0}.
%%%%%%%%%%%%%%%%%%%%%%%%%%%%%%%%%      
%%%%%%%%%%%%%%%%%%%%%%%%%%%%%%%%%
According to it, we can examine Fig.~\ref{ennebeta} assuming to be
viable only those models which, at the 1--$\sigma$ level, admit $n
>\sim 0.85\, $. With reference to this admittedly qualitative
criterion, we shall now consider the variable coupling case.

\begin{figure}
\begin{center}
\includegraphics*[width=15cm]{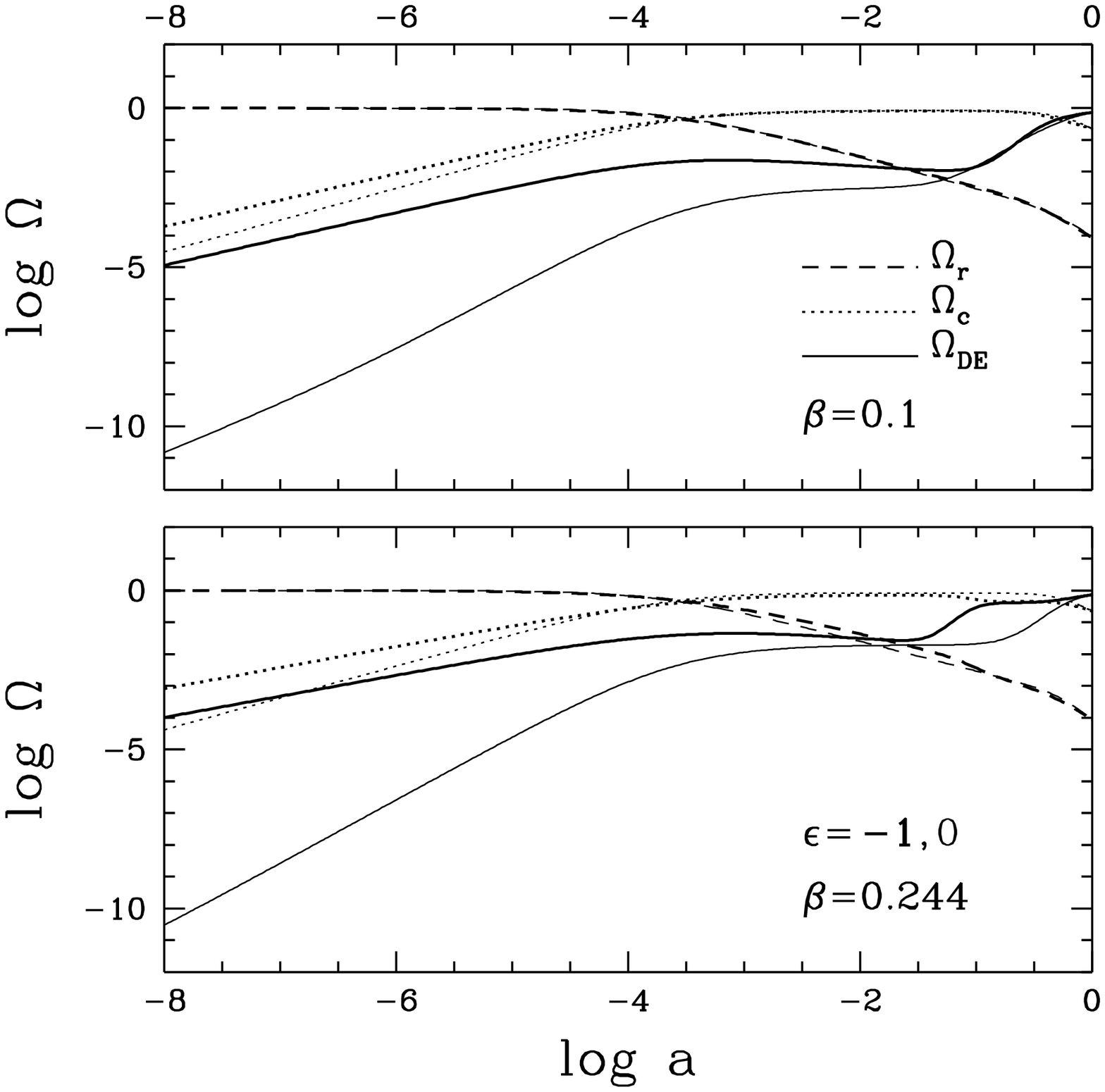}
\end{center}
\vskip -1.2truecm
\caption{Redshift dependence of density parameters $\Omega$ in coupled
DE models with constant and variable coupling. The two panels refer to
different values of $\beta$. In both panels we show $\Omega$'s for
$\epsilon=0$ ($C \propto \beta/m_p$, thinner lines) and $\epsilon=-1$
($C \propto \beta/\phi$, thicker lines). Figures are rather intricate
and their reading may begin from dashed lines, yielding $\Omega_r =
\Omega_\gamma + \Omega_\nu ,$ which are almost independent from the
coupling law. A more relevant effect occurs on $\Omega_c$ (dotted
lines), whose values, in the case of variable coupling, exceed those
of constant coupling until equality. The most relevant effect occurs
for DE (solid lines), whose contribution to the overall density is
enhanced by several order of magnitude by variable coupling. DE
fluctuations will fade after the entry in the horizon; however, high
$\Omega_{DE}$ values increase their impact on other components before
their disappearance.  }
\label{omega}
%\vskip -.4truecm
\end{figure}

First of all, when coupling varies, the evolution of the density
parameters exhibits significant discrepancies from the constant
coupling behavior. The point is that, at high redshift, they further
strengthen DM self--gravity, coherently with higher $\tilde \beta$
effects. Results of a numerical integrations, illustrating this issue,
are shown in Figure \ref{omega}, where we compare the redshift
dependence of density parameters in constant ($\epsilon = 0$) and
strongly variable ($\epsilon = -1$) coupling models, allowing to
appreciate a substantial enhancement of $\Omega_c$ at the eve of
equality. Altogether $[1+ 4 \tilde \beta^2(\phi)/3] \Omega_c $ is
significantly greater, and the freeze of $\delta_c,$ between horizon
entry and equality, is almost canceled.

%%%%%%%%%%%%%%%%%%%%%%%%%%%%%%%%%      
%%%%%%%%%%%%%%%%%%%%%%%%%%%%%%%%%
%%%%%%%%%%%%%%%%%%%%%%%%%%%%%%%%%      
\begin{figure}
\begin{center}
\includegraphics*[width=20cm]{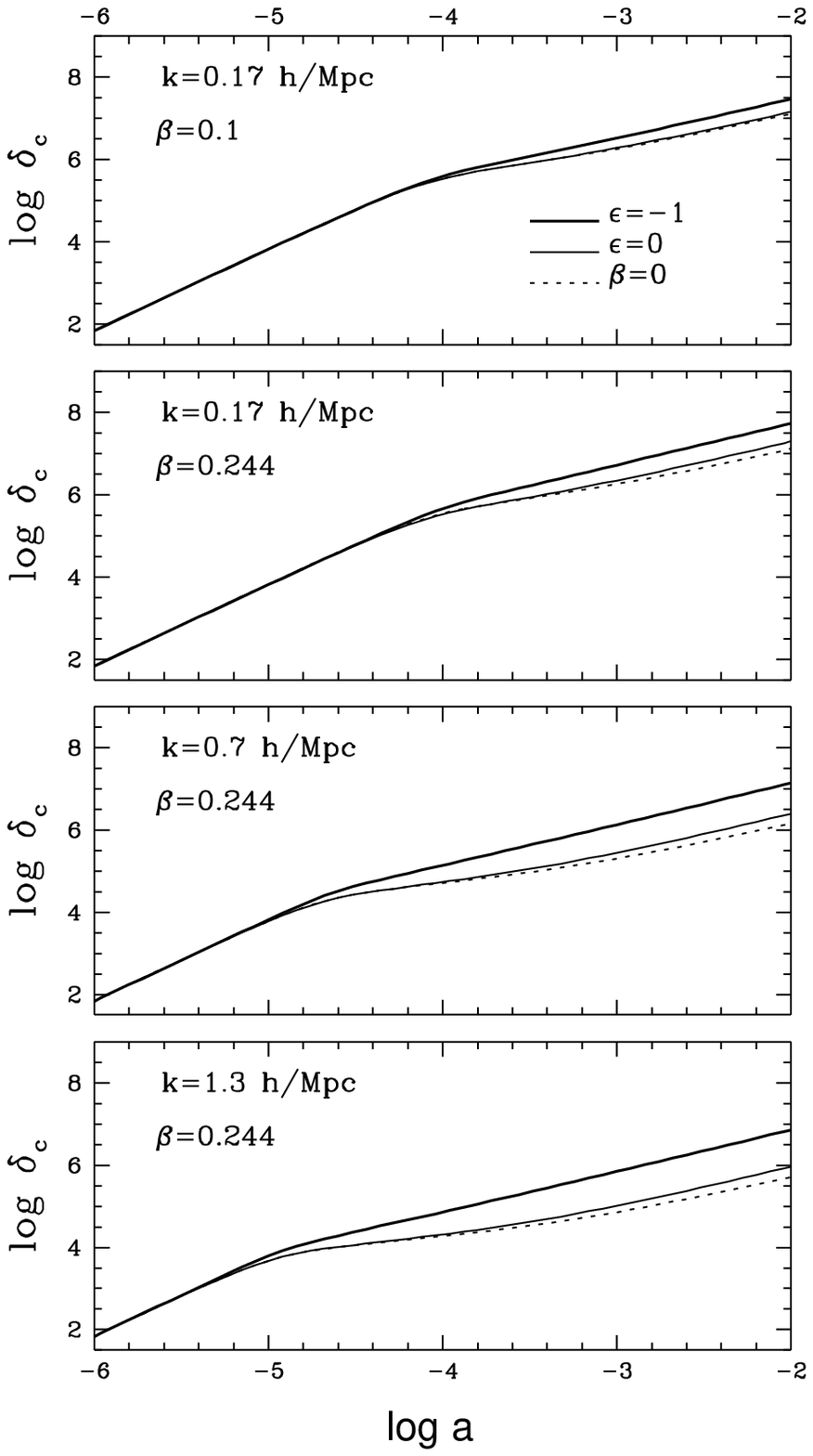}
\end{center}
\vskip -2.truecm
\caption{Evolution of DM fluctuations in a time interval enclosing
the entry in the horizon and the matter--radiation equality.
In all cases, in the presence of DM--DE coupling some modification
occurs. They are however almost negligible for constant coupling,
while, for coupling $\propto \phi^{-1}$, Meszaros' freezing is
almost completely canceled.
 }
\label{delta}
%\vskip -.4truecm
\end{figure}
%%%%%%%%%%%%%%%%%%%%%%%%%%%%%%%%%

This explains the behaviors shown in Figure \ref{delta}, which exhibit
one of the main findings of this work. These plots are obtained from a
numerical evaluations of $\delta_c,$ in models with different $\beta$
for $\epsilon = 0$ or -1. In the former case, the high--$z$ effect of
coupling is marginal. In the latter one, the fluctuation growth is
substantially enhanced, more significantly for greater $k$ values.

Accordingly, we can expect modifications in the transfer function and
in the fit of observational data. Our general conclusion is that a
time dependence of DM--DE coupling, making it stronger at higher
redshift, can prevent DM fluctuations to have a stationarity period
after their entry in the particle horizon, so causing large
modification of the transfer function.

\section{Results}
%%%%%%%%%%%%%%%%%%%%%%%%%%%%%%%%%      
\begin{figure}
\begin{center}
\includegraphics*[width=15cm]{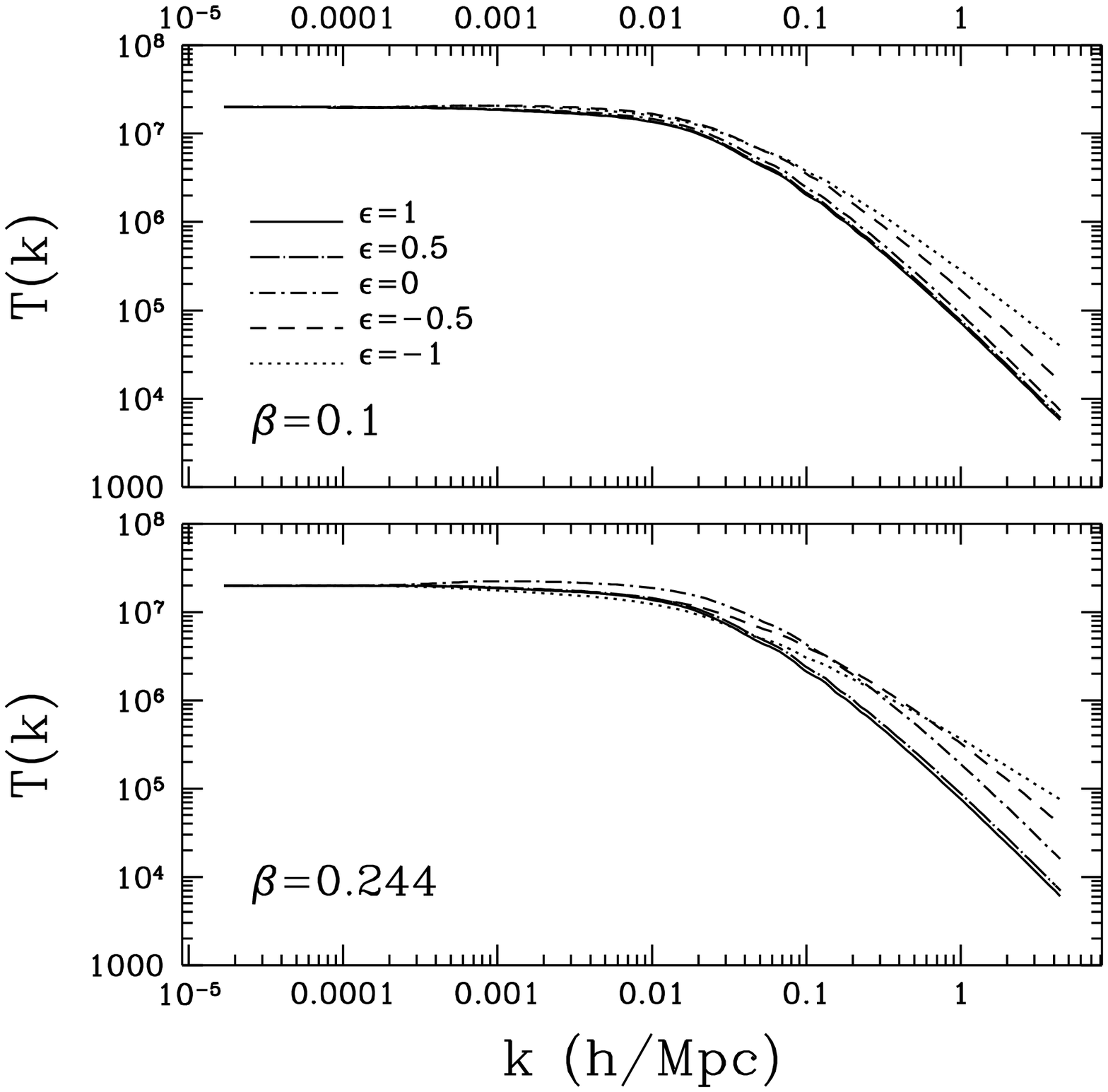}
\end{center}
\vskip -1.2truecm
\caption{Transfer functions for different behaviors of DM--DE coupling
with redshift and/or for different coupling normalization. The case
$\epsilon = 0$ corresponds to redshift independent coupling
intensity. The case $\epsilon = -1$ with $\beta = 0.244$ correspond to
a coupling $C = 1/\phi\, $.  Besides of the different slopes, notice
the dependence on the model of the bending scale and, in particular,
its dependence on the coupling strength, also in constant coupling
models (dash--dotted lines).  }
\label{transfer}
%\vskip -.4truecm
\end{figure}
%%%%%%%%%%%%%%%%%%%%%%%%%%%%%%%%%
These expectations will be tested by using an algorithm solving the
whole set of dynamical equations. To this aim we can use our extension
of the public programs CAMB, or our own code, with identical outputs.

Results for transfer functions are shown in Fig.~\ref{transfer} for
$\beta= 0.1$ and 0.244, and a variety of values~of~$\epsilon.$

The suppression of fluctuation freezing is obviously stronger for
greater $\beta $ (and increasingly negative $\epsilon$ values). For
$\epsilon = -1$, enclosing the case $C = 1/\phi$ when $\beta = 0.244$,
the steepness of the transfer function, for $k > k_{hor,eq}$ is much
reduced.  The effect is still significant also for $\epsilon = -0.5,$
namely when~$\beta = 0.244\, .$ 

A further effect shown by these plots is a significant displacement of
the scale where ${\cal T}$ begins its gradual descent. As a consequence, 
different coupling laws may cause displacements on the scale where
transfered spectra peak.

\begin{figure}
\begin{center}
\includegraphics*[width=15cm]{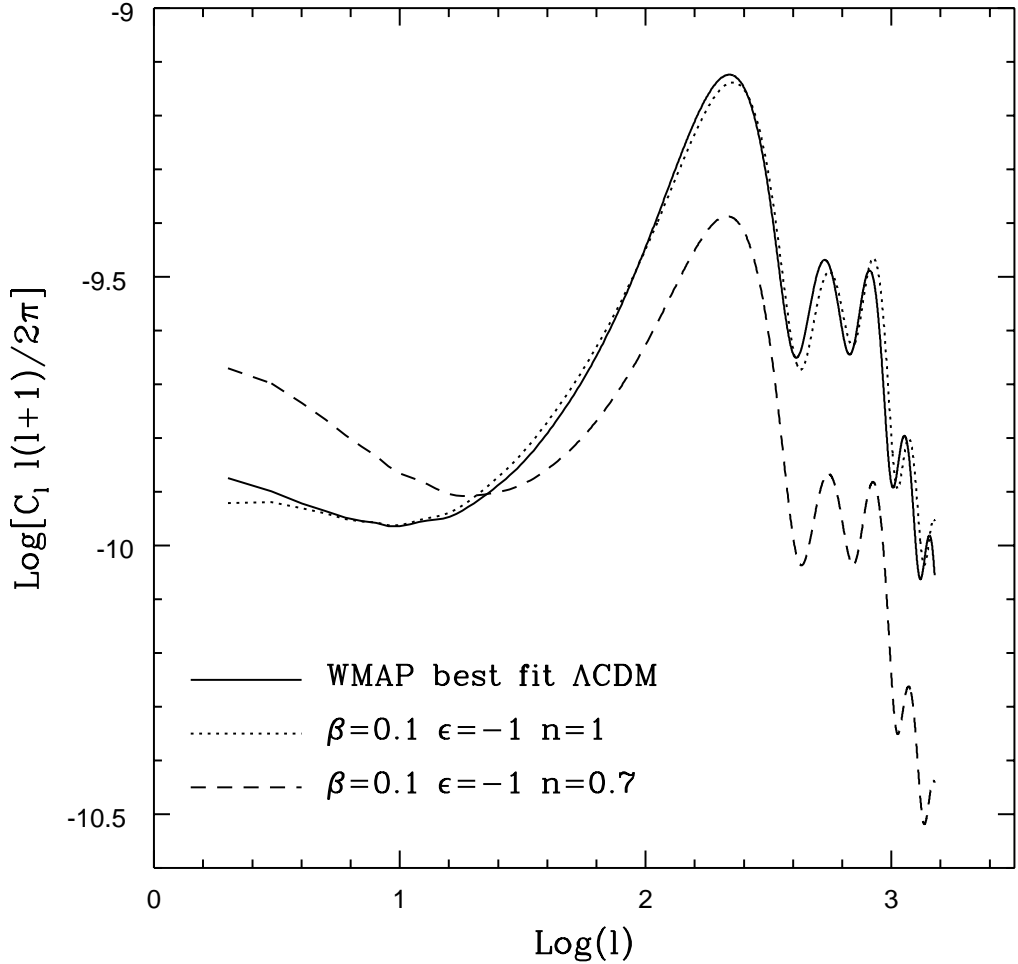}
\end{center}
\vskip -1.2truecm
\caption{Anisotropy spectrum of the $\Lambda$CDM model yielding the
best fit to WMAP3 data compared with the spectra for coupled models
with $\beta = 0.1$ and $\epsilon = 1,$ for $n = 1 $ and $n = 0.7.$ As
in Fig.~\ref{Cl0}, in the former case one can expect to recover a
reasonable fit to data by adjusting other model parameters . Taking $n
= 0.7,$ a value just acceptable to fit deep sample data, any fitting to CMB
anisotropy data is apparently excluded.  }
\label{Cl}
%\vskip -.4truecm
\end{figure}

Notice that, while this occur, the CMB anisotropy spectum keeps quite
a reasonable behavior, if $n \simeq 1,$ as is shown in
Figure~\ref{Cl}, similar to Fig.~\ref{Cl0} but for variable coupling;
here we compare the WMAP3 data on the anisotropy spectrum with the
spectra obtained for $\beta = 0.1$, $\epsilon = -1 $ and two $n$
values: $n = 1$ apparently allowing a reasonable fit, and $n = 0.7,$
as needed to obtain a reasonable transfered spectrum.  Figure
\ref{transfer}, in fact, is a direct evidence that, in order to
recover a fair slope of the transfered spectrum in the scale range
where structures accumulate, small primeval $n$ values are unavoidably
required. In order to perform an evaluation of the effect, we actually
built transfered spectra and compared them again with SDSS data. In
this way we find $n \sim 0.5$--0.7. Fig.~\ref{Cl} shows that this
spoils the fit with $C_l$ data.

Spectra with ordinary downward bending fit both deep sample and CMB
data with $n \simeq 1.$ On the contrary, spectra with low $n$'s and
standard $\sigma_8$'s cause greater $C_l$ still in the Sachs \& Wolfe
plateau, while reducing the relative height of the $C_l$ peaks (see
again Fig.~\ref{Cl}. These effects could be partially compensated by
an adjustment of other model parameters, whose search is out of the
scopes of this work. Finding $n$ values below 0.8--0.9, however,
clearly means that we are dealing with unlikely physical frameworks,
hardly allowing to fit CMB and deep sample data simultaneously.
%%%%%%%%%%%%%%%%%%%%%%%%%%%%%%%%%      
%%%%%%%%%%%%%%%%%%%%%%%%%%%%%%%%%

\begin{figure}
\begin{center}
\includegraphics*[width=11cm]{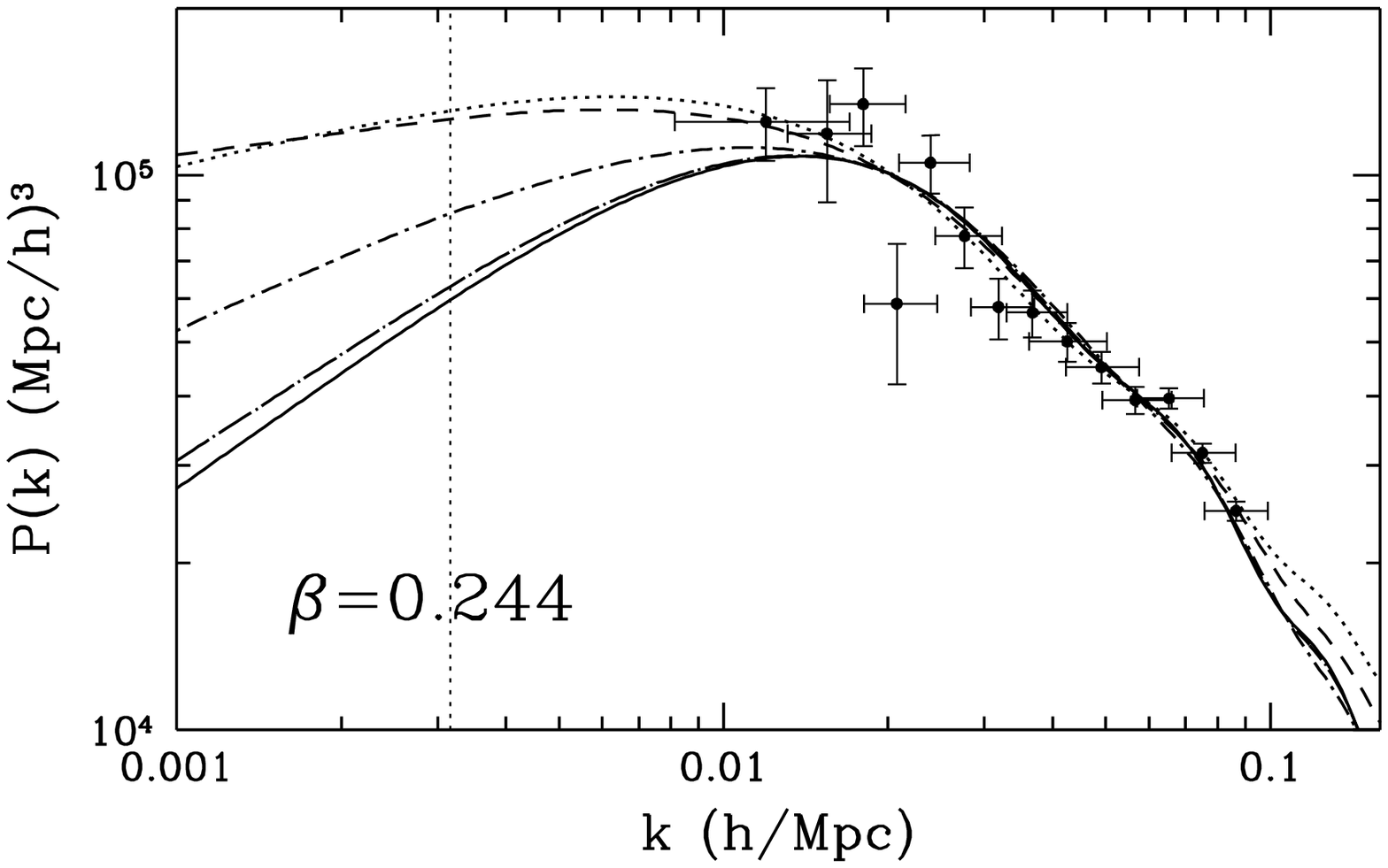}
\end{center}
\vskip -4.6truecm
\caption{Model comparison with SDSS digital survey data.  Different
curves refer to different values of $\epsilon $ (as in previous
Figure) with the solid line ($\epsilon = 1$) essentially coinciding
with an uncoupled model. Constant coupling models ($\epsilon = 0$) are
described by dot--dashed curves. Negative $\epsilon$'s yield a further
decrease of $n$. The vertical dotted line is the approximate scale
where the Sachs \& Wolfe $C_l$ plateau begins. Constant coupling
causes a rise of $C_{10}$ by a factor $\sim 1.8\, .$ A further factor
$\sim 2$ arises from a coupling $C = 1/\phi$.  }
\label{fitbeta2}
\vskip -1.2truecm
%\end{figure}
%%%%%%%%%%%%%%%%%%%%%%%%%%%%%%%%%
%%%%%%%%%%%%%%%%%%%%%%%%%%%%%%%%%      
%\begin{figure}
\begin{center}
\includegraphics*[width=11cm]{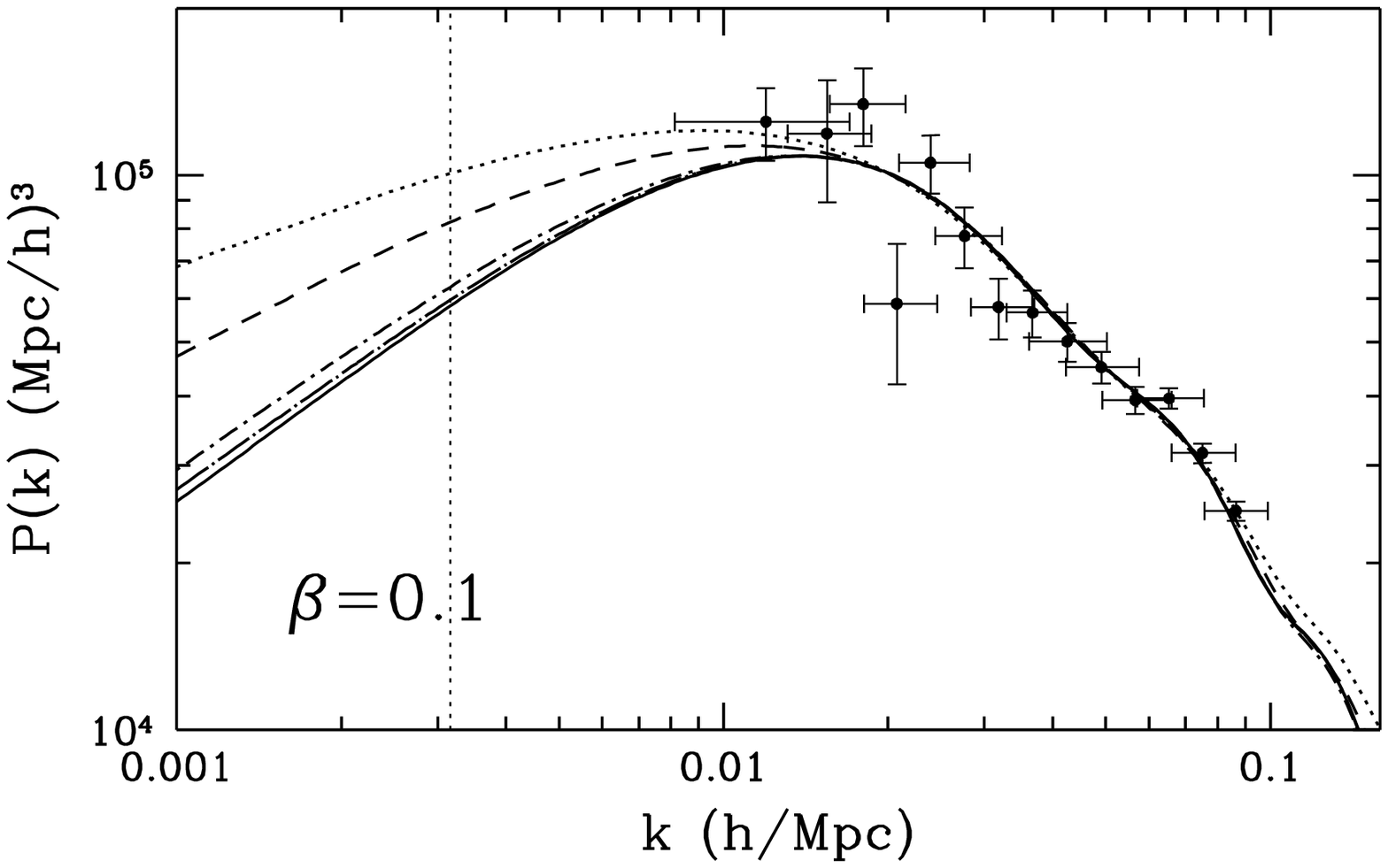}
\end{center}
\vskip -4.6truecm
\caption{As fig.~\ref{fitbeta2}, for a smaller coupling intensity.
The effect on the $C_{10}$ scale is much reduced, for constant
coupling; however, variable coupling yields a dramatic decrease
of $n$ and, for $\epsilon = -1,$ the level of $\beta = 0.244$
is almost attained.
}
\label{fitbeta1}
\vskip -.6truecm
\end{figure}

\begin{figure}
\begin{center}
\includegraphics*[width=11cm]{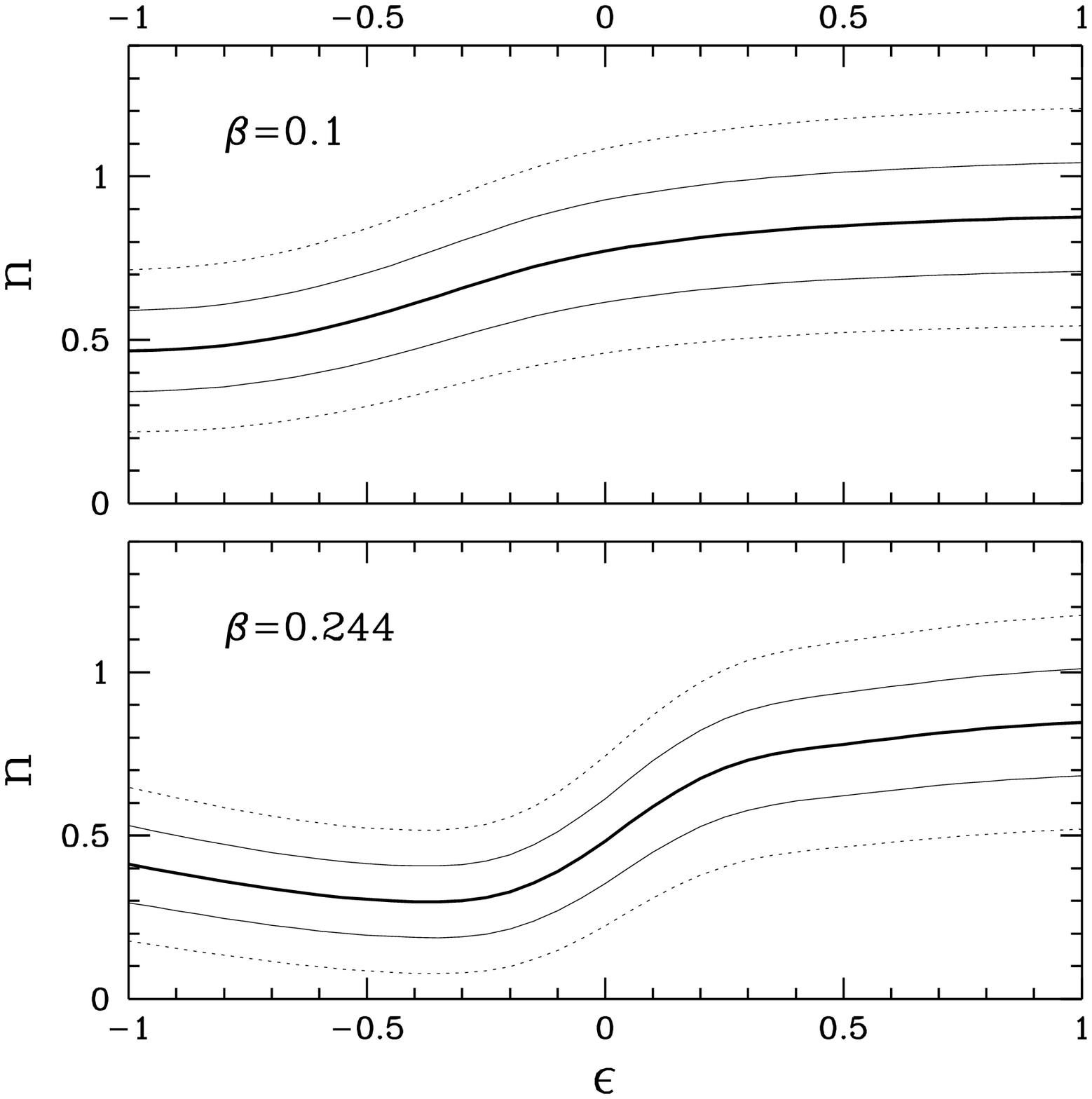}
\end{center}
\vskip -1.3truecm
\caption{1-- and 2--$\sigma$ intervals of $n$, when $\epsilon$ varies,
for $\beta$ values.}
\label{ennealfa}
%\vskip -.4truecm
%\end{figure}
%%%%%%%%%%%%%%%%%%%%%%%%%%%%%%%%%
%%%%%%%%%%%%%%%%%%%%%%%%%%%%%%%%%      
%\begin{figure}
\begin{center}
\includegraphics*[width=11cm]{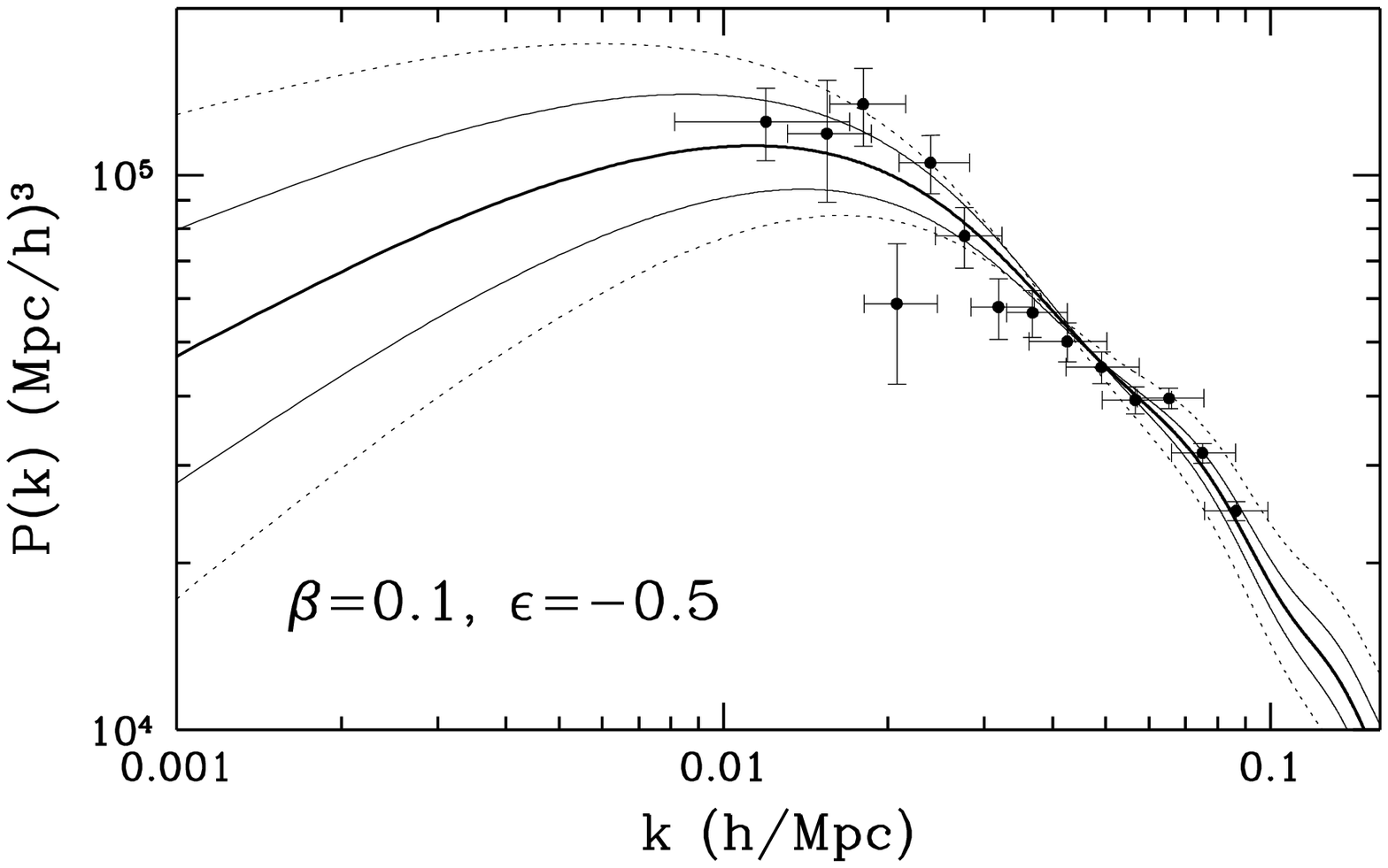}
\end{center}
\vskip -4.2truecm
\caption{If values of $n$ at 1-- or 2--$\sigma$'s from best fits are
taken, spectra are significantly modified. Here we show the effect in
the case with $C = 0.4(\pi/3)^{1/2}/\sqrt{m_p \phi}.$ }
\label{fitsigma1}
%\vskip -.4truecm
\end{figure}

In the case $\epsilon \neq 0,$ the discrepancy from unity of the
spectral index $n,$ assumed to be constant, is mostly a measure of the
distortion caused by the suppression of Meszaros effect, which
overwhelms the effects of the displacement of $k_{hor,eq},$ already
considered in the $\beta = {\rm const.}$ case. Moreover, such
discrepancy is a significant estimate of the distance of the model
from uncoupled physics. Our fits, shown in Figures \ref{fitbeta2}
and \ref{fitbeta1}, are complemented by 1-- and 2--$\sigma$ intervals
around best--fit $n$ values (Figure \ref{ennealfa}); they are an
indication of which models, by adjusting other parameters, might be
susceptible to approach the observational scenario.
%%%%%%%%%%%%%%%%%%%%%%%%%%%%%%%%%      
%%%%%%%%%%%%%%%%%%%%%%%%%%%%%%%%%
%%%%%%%%%%%%%%%%%%%%%%%%%%%%%%%%%      
%%%%%%%%%%%%%%%%%%%%%%%%%%%%%%%%%
The behavior of transfered spectra in respect to data, when taking $n
$ values at 1-- and 2--$\sigma$ from best fits is also shown in Figure
\ref{fitsigma1} (similar to Fig.~\ref{fitsigma2}).

If we take again $n=0.85$ at 1--$\sigma $ as a threshold to discard a
model, no $\epsilon < 0$ coupled model is allowed with $\beta =
0.244,$ while $\epsilon < -0.16$ are also inhibited with $\beta = 0.1\,
$. At 2--$\sigma$'s the situation is not much improved for $\beta =
0.244$, while lower values of $\epsilon$ are admitted for $\beta =
0.1\, $. In particular, a model with $C=1/\phi$, as the {\it
dual--axion} model, lays outside of the range indicated.

The analysis was extended here to models with positive $\epsilon,$ for
which coupling rises while $\phi$ increases. A large deal of these
models is apparently allowed. 

\section{Conclusions}
The quest for models fitting observational data and avoiding fine
tuning and coincidence suggests to test cosmologies where DM is
coupled to dynamical DE. It is then important to recognize that quite
a few models, with constant or variable DM--DE coupling, are
consistent with observational data, although their likelihood might be
slightly smaller than uncoupled models.

Constraints on these models were known to arise from a number of
linear and non--linear effects. The main linear anomaly is the
existence of a prolongated period, after matter radiation equality,
when the $\phi$--field energy affects the expansion rate (the
$\phi$--MD epoch). As a consequence, the comoving distance of the last
scattering band can be different from ordinary models.  In principle,
this can be compensated by varying the present value of the Hubble
parameter $H_o.$ There are however severe limits on $H_o,$ which turn
into limits on the coupling strength, discussed in \cite{cmb}. The same
feature causes also a displacement of the wave number $k_{hor,eq},$
corresponding to the scale which enters the horizon at
matter--radiation equality. Here we discussed also this effect, which
was however expected.

Constraints on coupled DE models were also found by studying their
non--linear evolution. In \cite{maccio} it is shown that coupling can
affect the halo concentration.  Although this effect could be softened
by using suitable self--interaction potentials, it yields a limit
$\beta < \sim 0.15$--0.20, for the coupling, when the
self--interaction potential is Ratra--Peebles \cite{RP}.

In this paper we outlined a new effect, that DM--DE coupling may cause
in a class of models. A prolongated period, between the horizon entry
and the equivalence, when DM fluctuation growth stagnates, is
essential in shaping transfered spectra $P_{tr}(k).$ Because of this
{\it Meszaros}'s effect, $P_{tr}(k)$ bends at $k > k_{hor,eq}.$ In
this paper we outline that this stagnation period can be partially or
totally suppressed by the coupling of DM with the $\phi$ field.  As a
result, transfered spectra exhibit a much softer bending at $k >
k_{hor,eq}.$ We also outline that, while this occurs, the dynamics of
sonic waves in the baryon--photon fluid is only marginally affected.
Accordingly, while transfered spectra suffer major changes, CMB
anisotropies are almost invariant.

Using a single primeval spectral index $n$, we can fit both CMB and
deep sample data, in the presence of standard Meszaros effect. In the
presence of the above softening, the values of $n$ required to fit CMB
and deep sample data become badly discrepant.

In this work we discuss which class of coupled DE models are affected
by this anomaly. If the coupling is constant or proportional to a
positive power of $\phi,$ the anomaly is absent. On the contrary,
models where the coupling is proportional to an inverse power of
$\phi$ are at risk.

We gauge the impact of this anomaly by evaluating the value of $n$
required to fit just deep sample data with each model.  When $n < \sim
0.85,$ it is legitimate to believe that the model likelihood is highly
suppressed. The search of such $n$ value should therefore be
preliminary to any attempt to reconcile a coupled model to the whole
set of CMB, deep sample (and other) data.

Admittedly, apart of conceptual reasons, within the present
observational framework there lacks any specific phenomenological push
to invoking a DM--DE coupling. Yet, uncoupled models, even apart of
their conceptual weakness, cause a number of questionable predictions,
e.g. NFW profiles. Interactions within the dark side were often
advocated to cure such difficulties.  An impact on profiles was
actually shown to exist, but new form of coupling need to be
inspected.

Furthermore, fresh data on the redshift dependence of $\rho_c$,
$\rho_b$ and $\rho_{DE}$, at $z \sim 1$--5, might soon be available,
if experiments like DUNE \cite{Dune} will become operational. The
discovery of an anomalous scaling of $\rho_c$, for instance, would set
a strong prior, completely biasing likelihood distributions, just as
priors on the value of $h$ (Hubble parameter) suppress the likelihood
of SCDM models with $h \sim 0.4\, ,$ which would otherwise allow a
reasonable fit of large sets of data \cite{Blanc}. In particular,
models with rising coupling ($\epsilon > 0$) could become an important
option. Such coupling behavior could directly arise from suitable
microphysics or be a phenomenological description of a complex
underlying physics.

\begin{ack}

Luca Amendola and Loris Colombo are gratefully thanked for their
comments on this work.

\end{ack}

\vfill\eject

{}

\end{document}